\documentclass[12pt, draftclsnofoot, peerreview, comsoc, onecolumn]{IEEEtran}
\usepackage[T1]{fontenc}

\ifCLASSINFOpdf
\else
\fi
\usepackage[T1]{fontenc}
\usepackage[noadjust]{cite}
\usepackage[T1]{fontenc}
\usepackage{textcomp}
\usepackage{amsmath}
\usepackage{xcolor}
\usepackage{soul}
\usepackage{amssymb}
\usepackage{algorithm}
\usepackage{algpseudocode}
\usepackage{subfigure}
\usepackage{graphicx}
\usepackage{epstopdf}
\usepackage{ulem}
\usepackage{caption}
\usepackage{pgfplots}
\pgfplotsset{compat=1.8}
\usepackage{tikz}
\usetikzlibrary{external}
\tikzset{external/force remake}
\usepgfplotslibrary{patchplots}
\usetikzlibrary{spy}
\usetikzlibrary{patterns}

\tikzset{
	font={\fontsize{10pt}{10}\selectfont}}

\pgfplotsset{compat=1.11,
	/pgfplots/ybar legend/.style={
		/pgfplots/legend image code/.code={%
			\draw[##1,/tikz/.cd,yshift=-0.25em]
			(0cm,0cm) rectangle (3pt,0.8em);},
	},
}

\pgfplotsset{compat=1.11,
	/pgfplots/xbar legend/.style={
		/pgfplots/legend image code/.code={%
			\draw[##1,/tikz/.cd,yshift=-0.15em]
			(0cm,0cm) rectangle (10pt,0.4em);},
	},
}
\begin{document}
\textcopyright 2018 IEEE. Personal use of this material is permitted. Permission from IEEE must be
obtained for all other uses, in any current or future media, including
reprinting/republishing this material for advertising or promotional purposes, creating new
collective works, for resale or redistribution to servers or lists, or reuse of any copyrighted component of this work in other works.
	\title{Ultra Reliable and Low Latency Communications in 5G {Downlink}: Physical Layer Aspects}
	
	\author{Hyoungju Ji$^{\dagger*}$,
		Sunho Park$^{\dagger*}$,
		Jeongho Yeo$^{*}$,
		Younsun Kim$^{*}$,
		Juho Lee$^{*}$,
		and Byonghyo Shim$^{\dagger}$
		\\$^{\dagger}$ Dept. of Electrical and Computer Engineering, Seoul National University, Korea
		\\$^{*}$ Samsung Electronics, Korea
	}

	\maketitle

	\begin{abstract}
  	   Ultra reliable and low latency communications (URLLC) is a new service category in 5G to accommodate emerging services and applications having stringent latency and reliability requirements. In order to support URLLC, there should be both evolutionary and revolutionary changes in the air interface named 5G new radio (NR). In this article, we provide an up-to-date overview of URLLC with an emphasis on the physical layer challenges and solutions in 5G NR downlink. We highlight key requirements of URLLC and then elaborate the physical layer issues and enabling technologies including packet and frame structure, scheduling schemes, and reliability improvement techniques, which have been discussed in the 3GPP Release 15 standardization.
 	\end{abstract}


	\section{Introduction}
    The new wave of the technology revolution, named the fourth industrial revolution, is changing the way we live, work, and communicate with each other. We are now witnessing the emergence of unprecedented services and applications such as driverless vehicles and drone-based deliveries, smart cities and factories, remote medical diagnosis and surgery, and artificial intelligence-based personalized assistants (see Fig. 1). Communication mechanisms associated with these new applications and services are way different from traditional human-centric communications in terms of  latency, energy efficiency, reliability, flexibility, and connection density. Therefore, coexistence of human-centric and machine-type services as well as hybrids of these will render emerging wireless environments more diverse and complex. To address diversified services and applications,  International Telecommunication Union (ITU) has classified 5G services into three categories: ultra-reliable and low latency communication (URLLC), massive machine-type communication (mMTC), and enhanced mobile broadband (eMBB) \cite{ITU-R}. To cope with these new service categories, various performance requirements such as massive connectivity, lower latency, higher reliability, and better energy efficiency have been newly introduced. Since the current radio access mechanism cannot support these relentless changes, 3rd Generation Partnership Project (3GPP) introduced a new air interface referred to as {\it New Radio} (NR) \cite{TR38913}. The primary goal of NR is to bring entirely new features and technologies that are not necessarily backward compatible with current 4G LTE systems. Currently, discussions for the 5G NR standardization are underway aiming at commercialization of the first release in 2020. 
		\begin{figure}[t!]
	\centering
	\includegraphics[width=150mm]{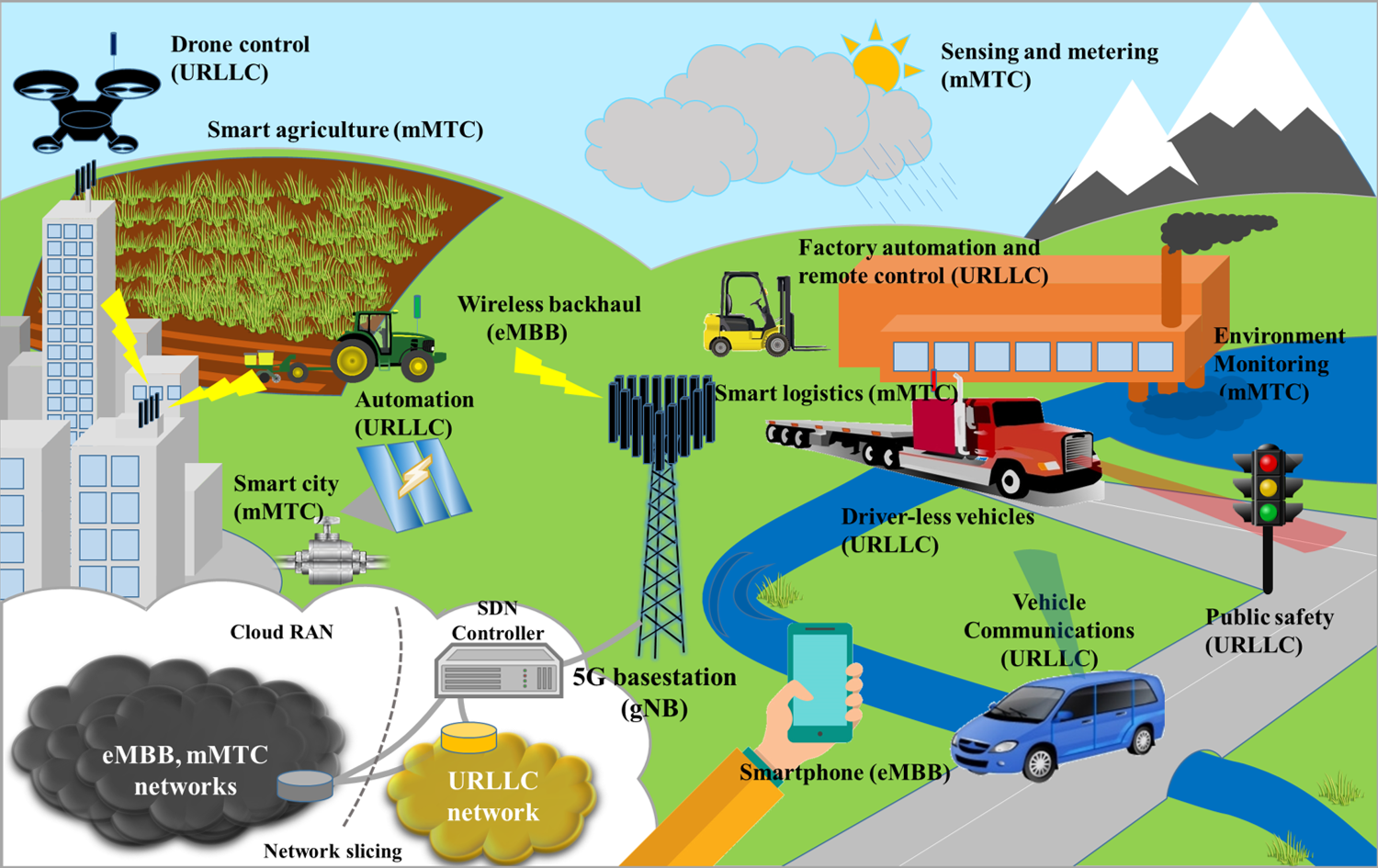}
	\caption{Overview of URLLC: deployment with URLLC and other services (eMBB and mMTC) under the network slicing.}
	\label{fig:URLLC1}
\end{figure}		
		
	Among the three categories we described, the design of URLLC is perhaps the most challenging. This is because URLLC needs to meet two challenging requirements: {\it low latency} and {\it ultra-high reliability}. When we try to improve the reliability, we need to use more resources for signaling, re-transmission, redundancy, and parity, resulting in an increase of the latency~\cite{RL}. In fact, in the current wireless systems whose sole purpose is to transmit a long packet to maximize the throughput, it is in general very difficult to achieve high reliability and low latency simultaneously. However, when the packet size is small, ultra reliability can be achieved at the expense of the achievable rate reduction \cite{cc}. This is because in many URLLC applications where the throughput requirement is not that stringent, reliability can be improved without violating the latency requirement by utilizing resources in the frequency, antenna, and spatial domain. Without doubt, interplay among throughput,  latency, and reliability requirements will make physical layer design of URLLC more complicated.

	The primary purpose of this paper is to present an up-to-date overview of the URLLC communications with an emphasis on the technical challenges and solutions in 3GPP NR downlink. We first describe the URLLC service requirements and then discuss physical layer issues and enabling technologies. These include packet and frame structure, scheduling schemes, and reliability improvement techniques. We also introduce early outcomes in 3GPP NR Release 15. 
	
	The rest of this paper is organized as follows. In Section II, we explain three service categories in 5G. In Section III, we discuss key requirements of URLLC. In Section IV, we present physical layer solutions for downlink URLLC services with simulation results and conclude the paper in Section V.
	
	\section{Three Service Categories in 5G}
    Before we proceed, we provide a brief overview of three service categories in 5G, viz., eMBB, mMTC, and URLLC.
    \subsection{eMBB}
    	eMBB is a service category related to high requirements for bandwidth, such as high-resolution video streaming, virtual reality, and augmented reality. The main challenge in 4G systems is to improve the system throughput (e.g., area, average, peak, perceived, and cell-edge throughput). Physical layer technologies introduced to this end include high order modulation transmission, carrier aggregation, cell densification via heterogeneous network, and  multiple-input multiple-output (MIMO) transmission. In essence, the main goal of eMBB is in line with this direction. In order to achieve 100-fold capacity increase over the 4G systems, more aggressive physical layer technologies improving the spectral efficiency and exploiting the unexplored spectrum are needed. Technologies under consideration  include full-dimension and massive MIMO \cite{Ji}, millimeter-wave communication \cite{mmW}, and spectrally-localized waveforms \cite{SI}. 
	\subsection{mMTC}
    mMTC is a service category to support the access of a large number of machine-type devices. mMTC-based services, such as sensing, tagging, metering, and monitoring, require high connection density and better energy efficiency \cite{noma}. Over the years, there have been some trials to support machine-type communications such as5 NB-IoT in licensed band, SigFox and LoRa in unlicensed band \cite{MIoT}. These approaches are similar in spirit but SigFox and LoRa technologies are suited for the stand alone services while NB-IoT is a good fit for the standard compatible services. These approaches offer some benefits, such as the low power consumption, low operation cost, and improved coverage. However, in the scenario where devices significantly outnumber the resources used for the transmission, an aggressive connection strategy violating the orthogonal transmission principle is required. In recent years, approaches using non-orthogonal spreading sequence or user-specific interleaving have been proposed to accommodate more users than the traditional approach relying on orthogonal multiple access \cite{noma}.
    
	\subsection{URLLC}
    URLLC is a service category to support the latency sensitive services such as remote control, autonomous driving, and tactile internet \cite{MIoT}. Since the time it takes for the human perception or reaction is in the order of tens of milliseconds (ms), packet transmission time for the mission-critical applications  needs to be in the order of tens$\sim$hundreds of microseconds (\textmu s) \cite{tactile}. While the latency of 4G LTE networks has been significantly improved from 3G networks, the end-to-end latency is still in the range of 30 $\sim$ 100 ms. This is because the backbone network typically uses the best-effort delivery mechanism and hence is not optimized for the mission-critical service. To reduce the end-to-end latency, therefore, there should be fundamental changes in both wireless link and backbone network. In the backbone link, software defined network (SDN) and virtual network slicing can be used to construct the private connection to the dedicated URLLC service \cite{MIoT}. Indeed, by using the dedicated network, latency of the backbone link can be reduced significantly. Whereas, in the wireless link, overhead should be reduced and the transmission mechanism needs to be streamlined. In fact, since a large portion of the transmit latency is due to the control signaling (e.g., grant and pilot signal) and it takes almost $0.3\sim0.4$ ms per scheduling, it is not so efficient to incorporate a low-latency packet transmission scheme in the current LTE systems. For example, when we design a short packet whose transmission latency is 0.5 ms, more than 60\% of resources would be wasted for the control overhead. To support URLLC, therefore, many parts of the physical layer should be re-designed.
	
	\section{URLLC Service Requirements}
	In order to come up with proper solutions to URLLC, it is necessary to understand the key requirements first. In this section, we present the latency and reliability requirements and then discuss the requirement related to the coexistence of URLLC and other services.
	 \begin{figure}[t!]
	 	\centering
	 	\includegraphics[width=135mm]{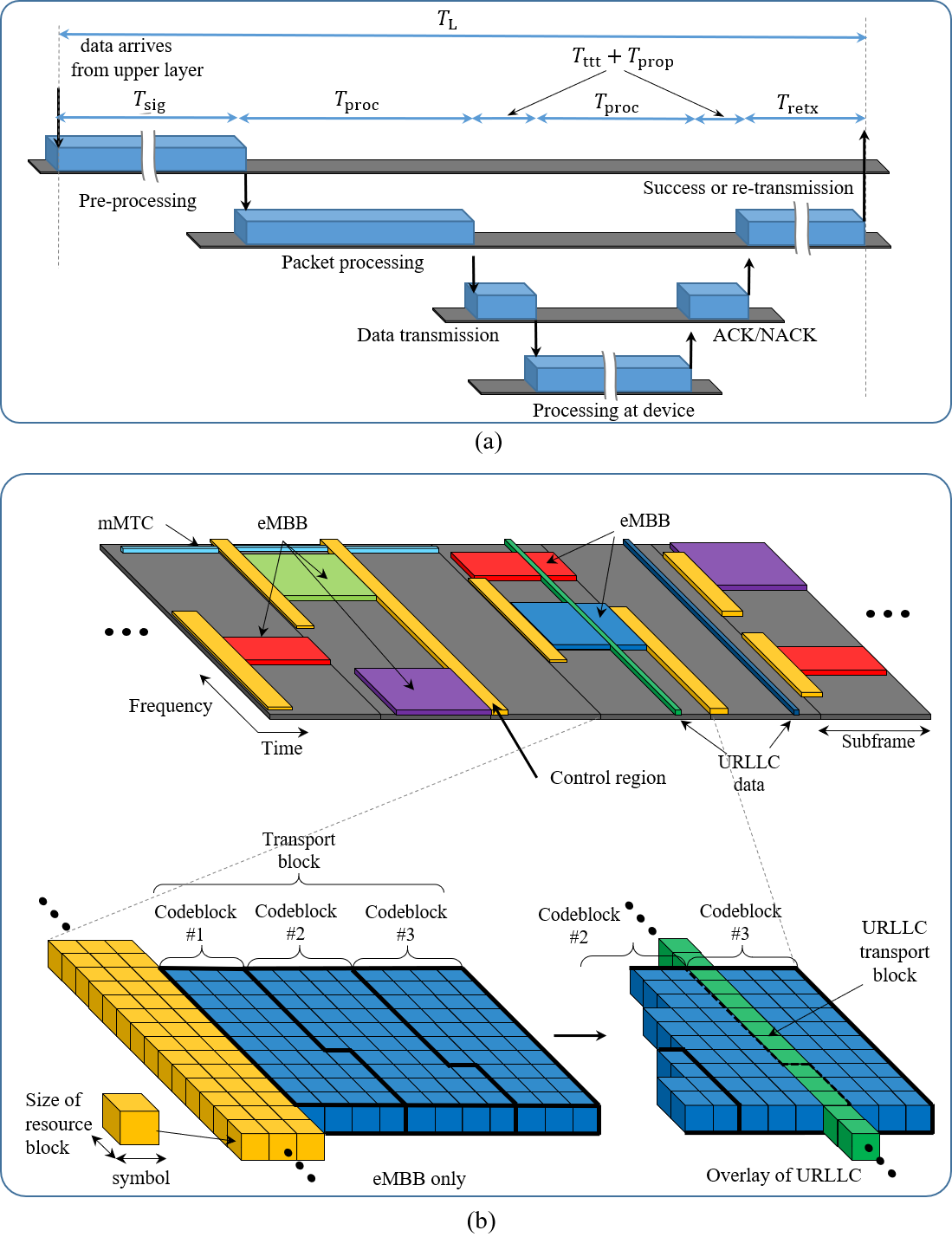}
	 	\caption{Physical layer downlink scenarios in URLLC service: a) illustration of latency components; b) transmission of  eMBB, mMTC, and URLLC packet in subframe level, and scheduling of URLLC packet into eMBB packet in symbol level.}
	 \end{figure}
	\subsection{Latency Requirement}
		Physical layer latency $T_\text{L}$ can be divided into the following five components  (see Fig. 2(a)):
		\begin{equation*}
		T_{\text{L}} = T_\text{ttt} + T_\text{prop} + T_\text{proc} + T_\text{retx} + T_\text{sig},
		\end{equation*}
		\begin{itemize}
			\item $T_\text{ttt}$  is the time-to-transmit latency which corresponds to the time to transmit a packet.
			\item $T_\text{prop}$ is the signal propagation time from the transmitter to the receiver.
			\item $T_\text{proc}$ is the time to perform the encoding and decoding, and also the channel estimation in the initial transmission.
			\item $T_\text{retx}$ is the time caused by the re-transmission.
			\item $T_\text{sig}$ is the pre-processing time for the signaling exchange such as connection request, scheduling grant, channel training and feedback, and queuing delay.
		\end{itemize}
        In response to the ITU requirements, 3GPP decided that the average latency of URLLC (from L2/L3 ingression to L2/L3 egression) should be less than $0.5$ ms  \cite{TR38913}. In order to meet this stringent latency constraint, a packet transmission time $T_\text{ttt}$ should be in the order of hundreds of microseconds. Since $T_\text{ttt}$ of the current 4G LTE systems is 1 ms period, a new frame structure reducing $T_\text{ttt}$ should be introduced. Also, since the latency caused by the channel estimation and feedback would be a bottleneck for the URLLC transmission, a transmission scheme that does not rely on the channel information needs to be considered.

	\subsection{Ultra-High Reliability}
 	  In 4G LTE systems, typical reliability for a packet transmission is $0.99$. Two key ingredients to achieve this goal are the channel coding (convolution and Turbo code) and the partial re-transmission of erroneous transport block called hybrid automatic repeat request (HARQ). URLLC services require much better performance, and in fact, the target reliability within 1 ms period should be at least $0.99999$ \cite{TR38913}. Further, in the mission-critical  applications such as autonomous driving and remote surgery, the reliability should be as high as $1-10^{-7}$ \cite{MIoT}. The first thing to do to meet these stringent requirements is to improve the  channel estimation accuracy. This is because the channel coding gain is small for the short packet so that the loss, if any, caused by the channel estimation should be prevented as much as possible. This is done by adding more resources to the pilot and using an advanced channel estimation technique. Even in this case, the required URLLC performance might not be satisfied so that additional resources in the frequency, antenna, and spatial domains are required to improve the reliability. Further, an advanced channel coding scheme suitable for the short packet transmission should be employed.\footnote{In LTE systems, 12 or 24 resource elements are allocated for demodulation reference signal (DMRS) per resource block (RB)}. In case the slot length is very short, repetitive transmission scheme using time-domain resources can also be a viable option.   
	       
	\subsection{Coexistence with eMBB and mMTC}	
 	 When there is a URLLC service request, whether in the scheduling period or in the middle of eMBB or mMTC transmission, the basestation should transmit the URLLC packet immediately~\cite{SI}. In other words, to support the URLLC packet transmission, ongoing eMBB and mMTC packets should be stopped without notice. As illustrated in the Fig. 2(b), when a transport block consisting of 3 codeblocks is transmitted for the eMBB service, each codeblock is mapped sequentially to the scheduled time-frequency resources. Thus, when the URLLC service is initiated in the middle of the eMBB transport block, part of symbols in the third codeblock  are replaced by the symbols of the URLLC packet. Since this interrupt is not reported to the mobile devices in use, reception quality of the eMBB and mMTC services will be degraded severely. This problem, dubbed as a coexistence problem in the 3GPP NR discussion, is a serious concern to the non-URLLC services so that a proper mechanism to protect the ongoing services should be introduced.
 	
	\section{URLLC Physical Layer in 5G NR}
 	In contrast to the 4G LTE systems, latency, reliability, and throughput requirements should be jointly considered in 5G NR so that there should be a fundamental change in the physical layer architecture (packet, slot, and frame). Specifically, a latency-sensitive packet structure for the fast decoding process and a flexible frame structure to support the dynamic change of the resource grid based on the latency requirement are needed. Also, when the URLLC  service is initiated, the URLLC packet should be transmitted instantly without delay. To do so, a scheduling scheme minimizing the transmit latency of the URLLC packet should be introduced. Further, since the latency requirement might not be satisfied by the HARQ-based re-transmission unless TTI of a packet is very short, a mechanism that significantly reduces the re-transmission latency is required.  
	Besides, an approach to use multiple radio interfaces to reduce the latency can be employed. Basic idea of this approach is to choose the radio access technology (RAT) providing the minimum latency among all possible options including 4G LTE, 5G NR, WiFi, and other IEEE 802.x standards. Using this together with the device-to-device (D2D) communications, the network layer latency can be reduced substantially.
		 	
	In this section, we put our emphasis on the physical layer solutions for URLLC including packet and frame structure to minimize the latency, multiplexing schemes to overlay the URLLC service into eMBB and mMTC services, and approaches to deal with the coexistence problem. We note that the reliability improvement and the latency reduction are equally important for the success of URLLC. However, there has been a consensus in the 3GPP NR standard meeting that the latency reduction issue should be considered by priority. This is because the reliability improvement can be achieved by the elaboration of 4G techniques such as channel coding, antenna, space, and frequency diversity schemes but such is not the case for the latency reduction effort \cite{Div, Coding}.

	\subsection{Packet Structure}
   The key issue in the URLLC packet  design is to minimize the processing latency $T_\text{proc}$ and the time-to-transmit latency $T_\text{ttt}$. Note that $T_\text{proc}$ consists of the time to receive packets, acquire channel information, extract control (scheduling) information, decode data packets, and check errors. In LTE systems, a square-shaped packet structure is popularly used for the efficient utilization of the spectrum under the channel fading. Whereas, in 5G NR systems, a non-square packet stretched in the frequency axis is used as a baseline since this structure minimizes the transmission latency $T_\text{ttt}$ \cite{38211}. Furthermore, in order to reduce the latency $T_\text{proc}$, three components of a packet, (pilot, control, and data part) should be grouped together to make a pipelined processing of the channel acquisition, control channel decoding, and data detection (see Fig. 3(a)).

	Another important issue to be considered is to use an advanced channel coding scheme. In 4G systems, two types of approaches have been employed to ensure reliability requirement. The first type is the channel coding scheme (Turbo and convolution code) with cyclic redundancy check (CRC) attachment for the large-sized packet. The second type is to use a simple code (the repetition and  Reed-Muller code) without CRC attachment for the small-sized packet. 
    In 5G NR, Polar code and low density parity check (LDPC) code have been adopted for the enhancement of control and data channel, respectively.
    Over the years, many efforts have been made to improve the decoding performance and computational complexity (and hence processing latency) of these codes such as successive cancellation list decoding of Polar code and non-binary LDPC decoding~\cite{Coding}.

    A recently proposed approach for the second type is the sparse vector coding (SVC), a short-packet transmission scheme based on the principle of compressed sensing (CS) \cite{CCSE}. Basic idea of SVC is to map the information into the position of a sparse vector $\mathbf{s}$. Note that when we choose $k$ positions in the $n$-dimensional vector, $\lfloor \log_2 \binom {n}{k} \rfloor$-bit information can be encoded. For example, if $k=2$ and $n=9$, the 5-bit of information can be encoded (e.g., $\mathbf{s} = [{\footnotesize 0 \ 1 \ 0 \ 0 \ 0 \ 0 \ 1 \ 0 \ 0}]^T$). By spreading the sparse vector $\mathbf{s}$ using the non-orthogonal spreading sequences $\mathbf{c}_i,~i=1,\dots,n$, we obtain the transmit vector $\mathbf{x}=[\mathbf{c}_1~\mathbf{c}_2~\cdots~\mathbf{c}_n]\mathbf{s}$ whose dimension $m$ is smaller than $n$ (the dimension of $\mathbf{s}$). 
	Since $\mathbf{s}$ is a sparse vector and $k$ is known in advance, the information vector $\mathbf{s}$ can be recovered via the CS technique \cite{tip}. A well-known advantage of the CS technique is that an input vector $\mathbf{s}$ can be recovered using a small number of measurements (resources). Since the computational complexity of the sparse recovery algorithm is proportional to the sparsity $k$, latency caused by the CS-based algorithm would be very marginal. 
	As shown in Fig. 4(a), physical downlink control channel (PDCCH) with SVC is effective in the short packet transmission and outperforms PDCCH using the convolution code in LTE-Advanced systems (3.1 dB at $10^{-5}$ PER).

\begin{figure}[t!]
	\centering
	\includegraphics[width=160mm]{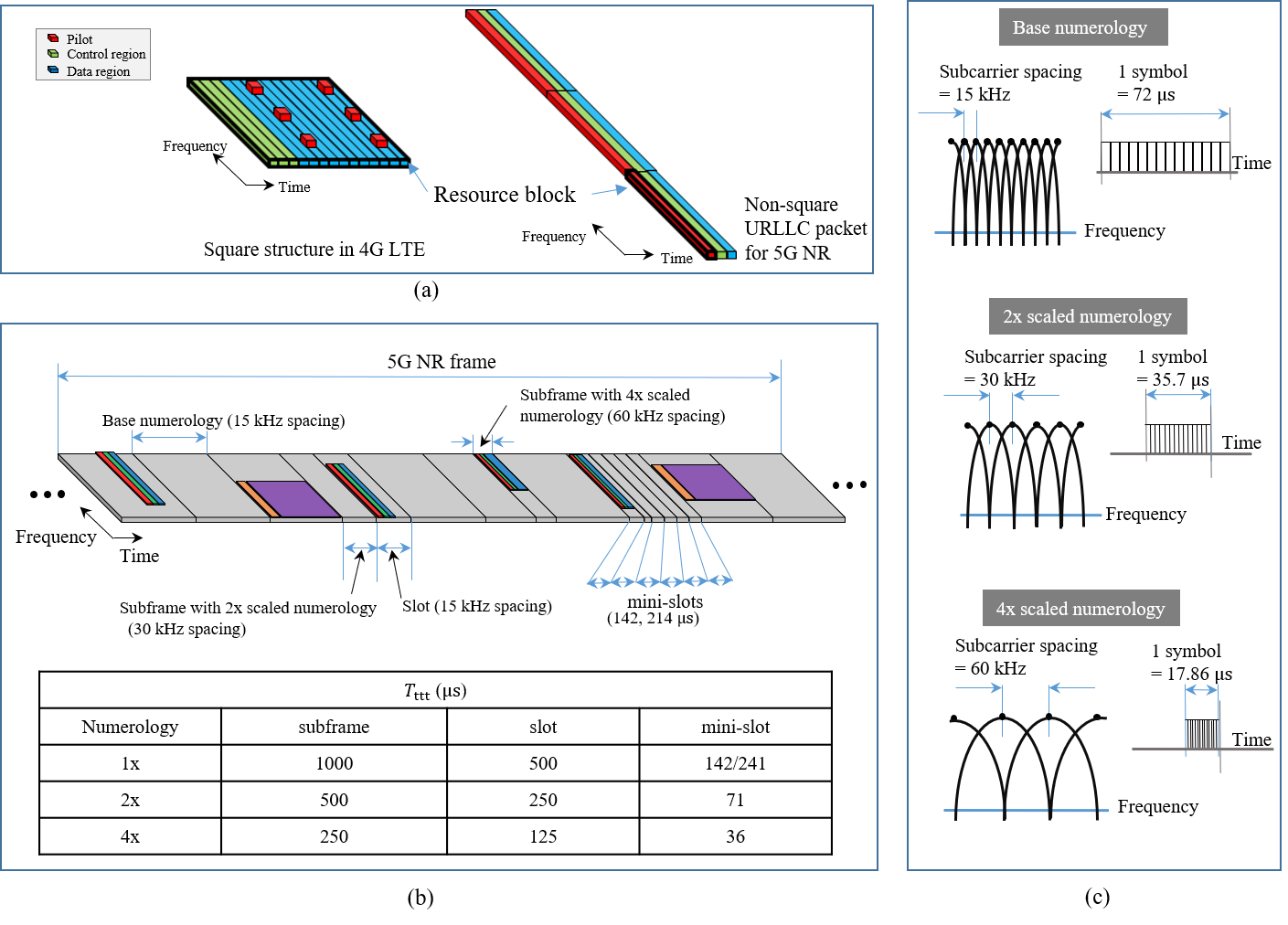}
	\caption{Packet and frame structure for URLLC: (a) packet structure; (b) frame structure; (c) supported numerologies for 5G NR.}
\end{figure}

	\subsection{Frame Structure and Latency-sensitive Scheduling Schemes}
	One of the main goals in 5G NR is to design a unified frame structure to cover a wide-range of frequency band and various service categories. To this end, flexible frame structure and user scheduling mechanism have been introduced.
	\subsubsection{Flexible Frame for URLLC}
	One direct option to reduce the time-to-transmit latency $T_\text{ttt}$ is to reduce the symbol period (see Fig. 3(b)). When the frequency band above 6 GHz (millimeter-wave) is used, due to the path loss, cell radius would be much smaller than that of conventional cellular systems and so will be the channel delay spread. In this case, by controlling the subcarrier spacing, we can reduce the symbol period  (see Fig 3(c)). For instance, the symbol length can be reduced by half (from 72 \textmu s to 36 \textmu s) by doubling the subcarrier spacing (from 15 kHz to 30 kHz). In doing so, the time to transmit one subframe can be reduced by half (from 1 ms to 0.5 ms).  However, when the frequency band below 6 GHz is used, this option might not be desirable due to the large delay spread. In this case, one can alternatively consider reducing TTI of the packet. For example, using mini-slot level (2$\sim$3 symbols) and slot level (7 symbols) transmission, $T_\text{ttt}$ can be reduced to 142, 241, and 500 \textmu s, respectively. In short, by controlling the symbol period and also the number of symbols in a packet, $T_\text{ttt}$ being smaller than 1 ms can be achieved (see Table in Fig. 3(b)). Note that to support this flexible frame structure, an advanced receiver equipped with fast tracking, quick synchronization, and simultaneous decoding functions is needed.

	\subsubsection{{Scheduling} Schemes}
	Since the URLLC packet is generated abruptly, how to schedule this into existing services is an important issue in the system design. Two schemes adopted in 3GPP NR standard are the instant scheduling and reservation-based scheduling \cite{38211}. 
	\begin{itemize}
		\item Instant scheduling: Any ongoing data transmission is interrupted to initiate the URLLC packet. This protocol is effective in reducing the URLLC access time but causes a severe performance degradation. Therefore, an approach to mitigate the performance degradation of ongoing services is needed (see Section IV. C).
	 	\item Reservation-based scheduling: In this scheme, URLLC resources are reserved prior to the data scheduling. Two types of reservation schemes are {\it semi-static} and {\it dynamic} reservations. In the semi-static reservation scheme, the basestation infrequently broadcasts the configuration of the frame structure such as frequency numerology and service period. Whereas, in the dynamic reservation scheme, information on the URLLC resource is updated frequently using the control channel of a scheduled user. For example, if an eMBB packet consists of 14 symbols, then 10 symbols are used for the eMBB transmission and the rest are reserved for the URLLC service. Drawback of this approach is that when there is no URLLC transmission in the scheduled period, resources reserved for the URLLC service will be wasted. When compared to the semi-static reservation, the dynamic reservation requires additional control overhead to indicate the reservation information. Also, an overhead to ensure the reliability of the control signaling itself is unavoidable.
	\end{itemize}
	In Fig. 4(b), we present the simulation results of average latency of URLLC and average throughput of eMBB service. In each scheduling period, the packet size is chosen such that the target packet error rates (PER) of URLLC of $10^{-5}$. In the simulation, latency is defined as the time that a packet arrived at the scheduler is successfully delivered to the mobile. Thus, one can deduce that the throughput is degraded when the latency increases. From Fig. 4(b), we see that the instant scheduling strategy outperforms the reservation-based scheduling in terms of the average latency but causes a throughput loss of the eMBB service. We also observe that the dynamic reservation outperforms the semi-static reservation in terms of the latency due to the fast resource adaptation. 
   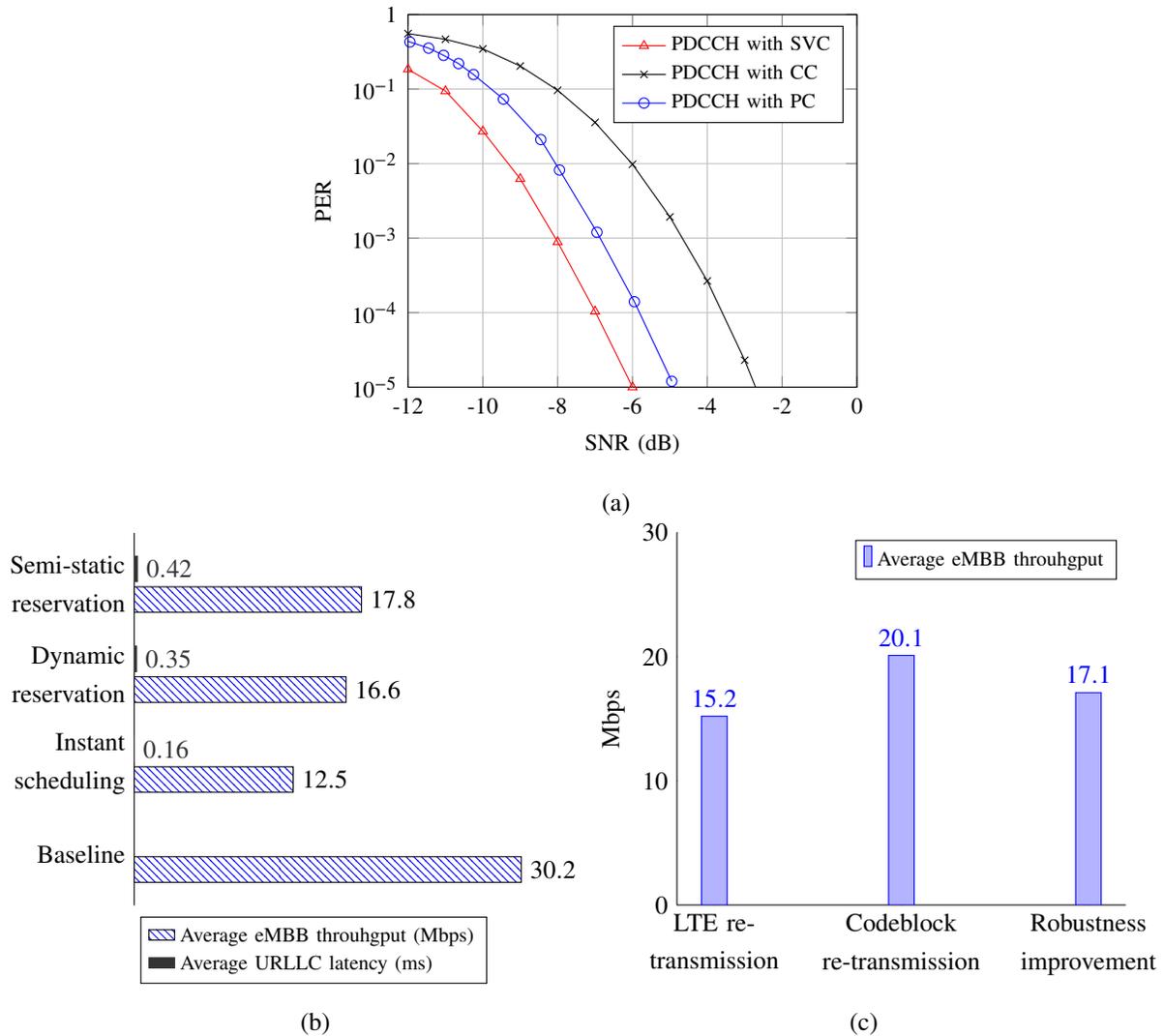
\begin{figure}[t!]
	\centering
\begin{tikzpicture}
\begin{semilogyaxis}[
xlabel=\footnotesize{SNR (dB)},
ylabel=\footnotesize{PER},
xmin=-12, xmax=0,
ymin=10e-6, ymax=1,
grid={major},
xtick={-12, -10,   -8  , -6, -4, -2, 0},
xticklabels={\footnotesize{-12}, \footnotesize{-10}, \footnotesize{-8},  \footnotesize{-6} , \footnotesize{-4}, \footnotesize{-2} ,\footnotesize{0}},
ytick={10e-6, 10e-5, 10e-4, 10e-3, 10e-2,10e-1, 1},
yticklabels={\footnotesize{$10^{-5}$}, \footnotesize{$10^{-4}$}, \footnotesize{$10^{-3}$}, \footnotesize{$10^{-2}$}, \footnotesize{$10^{-1}$}, \footnotesize{$1$}},
legend cell align=left,
legend style={legend pos=north east},
scale=0.9
]
\addplot[color=red,mark=triangle, mark options={solid}] coordinates {
	(-13, 0.25432)
	(-12, 0.18432)
	(-11, 0.09432)
	(-10, 0.027173)
	(-9, 0.006239)
	(-8, 0.000889)
	(-7, 0.000104)
	(-6, 1e-05)
	(-5, 1e-06)
	(2, 2.12e-07)
};
\addplot[color=black,mark=x, mark options={solid}] coordinates {
	(-12, 0.5555)
	(-11, 0.46655)
	(-10, 0.34655)
	(-9, 0.204)
	(-8, 0.096503)
	(-7, 0.035716)
	(-6, 0.009792)
	(-5, 0.001926)
	(-4, 0.000266)
	(-3, 2.3e-05)
	(-2, 1.3e-06)
};
\addplot[color=blue,mark=o, mark options={solid}] coordinates {
	(-13.45, 0.684) 
	(-12.95, 0.597)	
	(-12.45, 0.523) 	
	(-11.95, 0.429)
	(-11.45, 0.355) 
	(-11.05, 0.284) 
	(-10.65, 0.22) 
	(-10.25, 0.1569)
	(-9.45, 0.0734) 
	(-8.45, 0.021)  
	(-7.95, 0.0082) 
	(-6.95, 0.0012) 
	(-5.95, 0.00014)
	(-4.95, 0.000012)
};
\legend{\scriptsize{PDCCH with SVC}, \scriptsize{PDCCH with CC},  \scriptsize{PDCCH with PC}}
\end{semilogyaxis}
\end{tikzpicture}

\small{~~~~~~(a)}

\begin{tikzpicture}
\begin{axis}[
scale=0.9,
xbar,
axis lines*=left,
axis x line       = none,
tickwidth         = 0pt,
enlarge y limits  = 0.19,
enlarge x limits  = 0,
symbolic y coords = {Baseline, Instant scheduling, Dynamic reservation, Semi-static reservation},
xmax=35,
ytick=data,
legend cell align={left},
yticklabels={
	Baseline,
	Instant\\scheduling,
	Dynamic\\reservation,
	Semi-static\\reservation
},
legend style={at={(0.4,-0.03)},
	anchor=north},
yticklabel style={align=right},
nodes near coords,
]
\addplot[pattern=north west lines, pattern color=blue] coordinates { (30.2,Baseline)         (12.5,Instant scheduling)
	(16.6,Dynamic reservation)  (17.8,Semi-static reservation) };
\addplot[fill, color=black!80] coordinates {        (0.16,Instant scheduling)
	(0.35,Dynamic reservation)   (0.42,Semi-static reservation)  };
\legend{\scriptsize{Average eMBB throuhgput (Mbps)}, \scriptsize{Average URLLC latency (ms)}}
\end{axis}
\end{tikzpicture}
\begin{tikzpicture}
\begin{axis}[
scale=0.9,
ybar,
axis lines*=left,
tickwidth         = 0pt,
enlarge y limits  = 0,
enlarge x limits  = 0.1,
symbolic x coords = {LTE re-transmission, Codeblock re-transmission, Robustness improvement},
ymin=0,
ymax=30,
xtick=data,
x tick label style={text width=2.2cm,align=center},
ylabel={Mbps},
nodes near coords,
]
\addplot coordinates {       (LTE re-transmission,15.2)
	(Codeblock re-transmission,20.1)  (Robustness improvement,17.1) };
\legend{\scriptsize{Average eMBB throuhgput}}
\end{axis}
\end{tikzpicture}

\small{(b)~~~~~~~~~~~~~~~~~~~~~~~~~~~~~~~~~~~~~~~~~~~~~~~~~~~~~~~~~~(c)}
	\caption{Reliability and latency performance for the downlink transmission: a) PER of the short packet transmission (12 bit) for URLLC; b) average latency of URLLC scheduling schemes and impact on eMBB average throughput; c) average throughput performance of coexistence options when the instant URLLC scheduling occurs.}
\end{figure}
	\subsection{Solutions to the Coexistence Problem}
	As mentioned, the holy grail of 5G NR is to support diverse service categories and thus how to mitigate the performance degradation of interrupted services is an important issue in the physical layer design. While the flexible frame structure may ease off this problem, due to the implementation complexity and randomness of URLLC packet arrival, a more deliberate solution is required in real deployment scenarios. Two approaches discussed in the 5G NR standard meetings are {\it reactive} and {\it proactive} strategies. 
	\subsubsection{Reactive strategy}
	The main idea is to give a priority to the URLLC packet while ensuring the reliability of the other channel interrupted by URLLC. Two approaches adopted in the 3GPP NR are as follows \cite{SI}.
	\begin{itemize}
		\item Preemption indicator transmission: In this approach, the basestation indicates which resources are used for the URLLC transmission. Recalling that the URLLC packet is stretched in the frequency axis (see Fig. 3(a)), URLLC transmission will interrupt the whole system bandwidth and thus degrade all data channels in use. To notify this event to the scheduled users, the basestation broadcasts a {\it preemption indicator} consisting of time and/or frequency information of the interruption. This indicator helps users identify the reason for packet errors and what part of the packet is safe from the interruption. 
		\item Re-transmission of selected codeblocks: When the ongoing service is interrupted by the URLLC transmission, part of the codebook affected by URLLC is re-transmitted. By transmitting {\it combining indicator} or {\it flush-out indicator}, the receiver can perform the soft symbol combining of the transmitted and re-transmitted codeblocks. One can further achieve better coding gain by lowering the code rate of the re-transmitted codeblock.
	\end{itemize}
	\subsubsection{Proactive strategy}
	If the URLLC transmission occurs frequently, the efficiency of the reactive approach will be reduced due to the frequent re-transmissions. The main idea of the proactive strategy  is to ensure the reliability of ongoing services while supporting the URLLC transmission. Specific schemes to support the proactive strategy include robustness improvement and service sharing. 
	\begin{itemize}
		\item Robustness improvement: To reduce the initial packet error of non-URLLC packets, the basestation intentionally lowers the code rate by adding extra parity bits or employing outer error correction code to the non-URLLC packets \cite{38211}. Since the URLLC data transmission interferes  non-URLLC packets, this approach can help reducing the packet error of non-URLLC packets.
		\item Resource sharing: This strategy supports the ongoing data channel and the URLLC data channel simultaneously. Multiple antenna or beam-domain techniques are employed for this purpose. Basically, the spatial layer (rank) of the channel is divided into two and then one part of the layers is used for eMBB  and the other is used for URLLC. If there is no extra spatial layer, then the power-domain non-orthogonal transmission can be applied~\cite{noma}.
	\end{itemize}
    In Fig. 4(c), we plot the average throughput of 4G LTE HARQ, reactive strategy (re-transmission of selected codeblock), and proactive strategy (robust channel coding). We first observe that the performance degradation caused by the instant access is mitigated by the coexistence solutions. We also observe from Fig. 4(c) that the re-transmission of selected codeblocks outperforms the LTE HARQ scheme by a large margin, achieving 32\% gain in throughput. Whereas, due to the substantial parity overhead, we observe that the robustness improvement scheme suffers from throughput loss (about 17\%) over the reactive strategy.

	\section{Conclusion}
	URLLC is one of the key services in 5G communications having wide applications including automated controls, tactile internet, remote operations, and intelligent transportation systems. Despite its importance, the physical layer technologies to seamlessly integrate URLLC into 5G NR are in its infancy. In this article, we have discussed the key requirements for URLLC and presented the physical layer enabling technologies. In order to satisfy stringent latency and reliability requirements, many parts of the physical layer should be re-designed, and the techniques presented in this paper can serve as a starting point. Other than the solutions we discussed, there are many interesting issues worth exploring such as the beamforming strategies for control and data part and the reconfigurable URLLC protocol. Also, study of advanced transceiver architecture to support dynamic numerology adaptation and simultaneous decoding is needed. In this article, we put our focus on the URLLC transmission in the downlink, but there are many open issues for the uplink direction such as one-shot access, active user detection, and grant-free transmission.

\newpage	
	\section*{Acknowledgment}
	This work was supported by the National Research Foundation of Korea (NRF) grant funded by the Korean government(MSIP)(2014R1A5A1011478) and `The Cross-Ministry Giga KOREA Project' grant funded by the Korea government(MSIT) (No. GK17P0501, Development of Ultra Low-Latency Radio Access Technologies for 5G URLLC Service).

\begin{table}[t]
	\begin{center}
		\caption{{Simulation assumptions}}
		\label{tab:simulation}
		\begin{tabular}{p{2.5cm}|p{12cm}} \hline
			\textbf{Parameter} & \textbf{Value and description} \\ \hline \hline
			Coding scheme (Fig. 4(a)) & PDCCH with SVC: $n=92$, $m=42$, $k=2$ \cite{CCSE}, PDCCH with Convolution code (CC):  $\frac{1}{3}$ rate, PDCCH with Polar code (PC): $\frac{1}{4}$ rate \cite{Coding}\\  \hline
					Simulation method & Link level with 3 km/h extended pedestrian-A (EPA) channel  \\ \hline
		System model & 20 MHz bandwidth with 3.5 GHz center frequency, 15 kHz spacing, and 1 slot = 1 ms\\ \hline
			Scheduler & The frequency selective (6 RBs) for eMBB and full-bandwidth (100 RBs) for URLLC. \\  \hline
			Receiver & Minimum mean square error (MMSE) using DMRS \\ \hline
			URLLC  &  {$\bullet$ Instant scheduling}: Randomly selects one symbol within a slot \\
			transmission & {$\bullet$ Reservation}: 4 symbols are reserved (fixed in semi-static and adapted in dynamic) \\ 
			 & $\bullet$ TTI= $\frac{1}{14}$ ms, and maximum number of re-transmission = 2 \\ \hline
					 eMBB & {$\bullet$ LTE re-transmission}: set target block error rate (BLER) = $10^{-2}$\\
					transmission & {$\bullet$ Re-transmission of codeblock}: set target BLER = $10^{-2}$, per codeblock re-transmission \\
					 & {$\bullet$ Robustness improvement}: set target BLER = $10^{-3}$, per transport block re-transmission \\ 
					 & $\bullet$ TTI= 1ms, and maximum number of re-transmission = 4 \\ \hline	 
		\end{tabular}
	\end{center}
\end{table}

	\ifCLASSOPTIONcaptionsoff
	\newpage
	\fi

\end{document}